\documentclass[11pt]{article}   
\usepackage{hyperref}
\pdfoutput=1
\usepackage{t1enc}
\usepackage[latin1]{inputenc}
\usepackage[english]{babel}
\usepackage{natbib}
\usepackage[pdftex]{graphicx}
\usepackage{fnbreak} 
\usepackage{url}
\usepackage{authblk}

\bibpunct{(}{)}{;}{a}{}{,}

\begin{document}

\title{Night-sky brightness monitoring in Hong Kong \\- a city-wide light pollution assessment}


\author[1]{Chun Shing Jason Pun\footnote{email: jcspun@hku.hk}}
\author[1]{Chu Wing So\footnote{email: socw@hku.hk}}
\affil[1]{Department of Physics, The University of Hong Kong, Pokfulam Road, Hong Kong, PR China}


\maketitle

\begin{abstract}
Results of the first comprehensive light pollution survey in Hong Kong are presented.  The night-sky brightness was measured and monitored around the city using a portable light sensing device called the Sky Quality Meter over a 15-month period beginning in March 2008. A total of 1,957 data sets were taken at 199 distinct locations, including urban and rural sites covering all 18 Administrative Districts of Hong Kong. The survey shows that the environmental light pollution problem in Hong Kong is severe~-- the urban night-skies (sky brightness at 15.0 mag arcsec$^{-2}$) are on average $\sim$~100 times brighter than at the darkest rural sites (20.1 mag arcsec$^{-2}$), indicating that the high lighting densities in the densely populated residential and commercial areas lead to light pollution. In the worst polluted urban location studied, the night-sky at 13.2 mag arcsec$^{-2}$ can be over 500 times brighter than the darkest sites in Hong Kong. The observed night-sky brightness is found to be affected by human factors such as land utilization and population density of the observation sites, together with meteorological and/or environmental factors. Moreover, earlier night-skies (at 9:30pm local time) are generally brighter than later time (at 11:30pm), which can be attributed to some public and commercial lightings being turned off later at night. On the other hand, no concrete relationship between the observed sky brightness and air pollutant concentrations could be established with the limited survey sampling. Results from this survey will serve as an important database for the public to assess whether new rules and regulations are necessary to control the use of outdoor lightings in Hong Kong. \
\end{abstract}

\section{Introduction}
\label{intro}
Outdoor lighting is an indispensable element of modern civilized societies for safety, recreation, and decorating purposes. However, poorly designed lighting systems and excessive illumination levels have led to a huge waste of energy and money and certain environmental consequences. Light pollution, according to the International Dark-Sky Association \citep{IDA:define}, refers to any "adverse effect of artificial light including sky glow, glare, light trespass, light clutter, decreased visibility at night, and energy waste." Common sources of light pollution include artificial outdoor lightings such as street lamps, neon signs, and illuminated signboards. 

The invention of electric lightings had no doubt altered the natural rhythms of day and night in the ecosystem. The lost of darkness due to light pollution has become a serious threat for several species and habitats. Excessive outdoor lightings may disorient physiological cycles and the movements of animals \citep{starlight:2007}. For example, it was estimated that millions of birds across the North America are killed every year by crashing into windows after attracted by light and by trying to navigate by artificial lights instead of natural directional cues such as stars \citep{ogden:1996}. In the worst scenario, light pollution would even influence the entire ecological balance of the local living environment \citep{michael:2007}. Some studies and news reported have shown that excessive illuminations (in either intensity or in variability) near people's living environment may influence their health \citep{davis:2001,blask:2005,stevens:2006} even though the full physiological impacts of long-term exposure to these lighting on humans have not been fully established.

The wasteful light emitted directly by artificial lighting or by reflections from the ground can be scattered by clouds, fog, and pollutants such as suspended particulates in the atmosphere, brightening the night-sky and reducing its brightness contrast \citep{benn:newastro}. One direct consequence of light pollution is the number of stars visible by unaided eyes decreased. Astronomers are therefore among the worst affected by the worldwide growing light pollution problem due to the diminishing accuracy of astronomical observations on the dim celestial objects under light-polluted skies \citep{garstang:wilson,massey:2000,richard:2010}. 
The importance of starry nights in human civilizations throughout history (for examples, myths and legends in the Greek mythology, and developments of Newton's theory of gravity) means the progressive degradation of the night-sky should also be regarded as a fundamental loss of a common and universal heritage of the mankind \citep{starlight:2007}. Light pollution should therefore be considered as an imminent environmental risk that must be addressed in the same fashion as other environmental pollution problems.

\begin{figure}
\includegraphics[width=8cm]{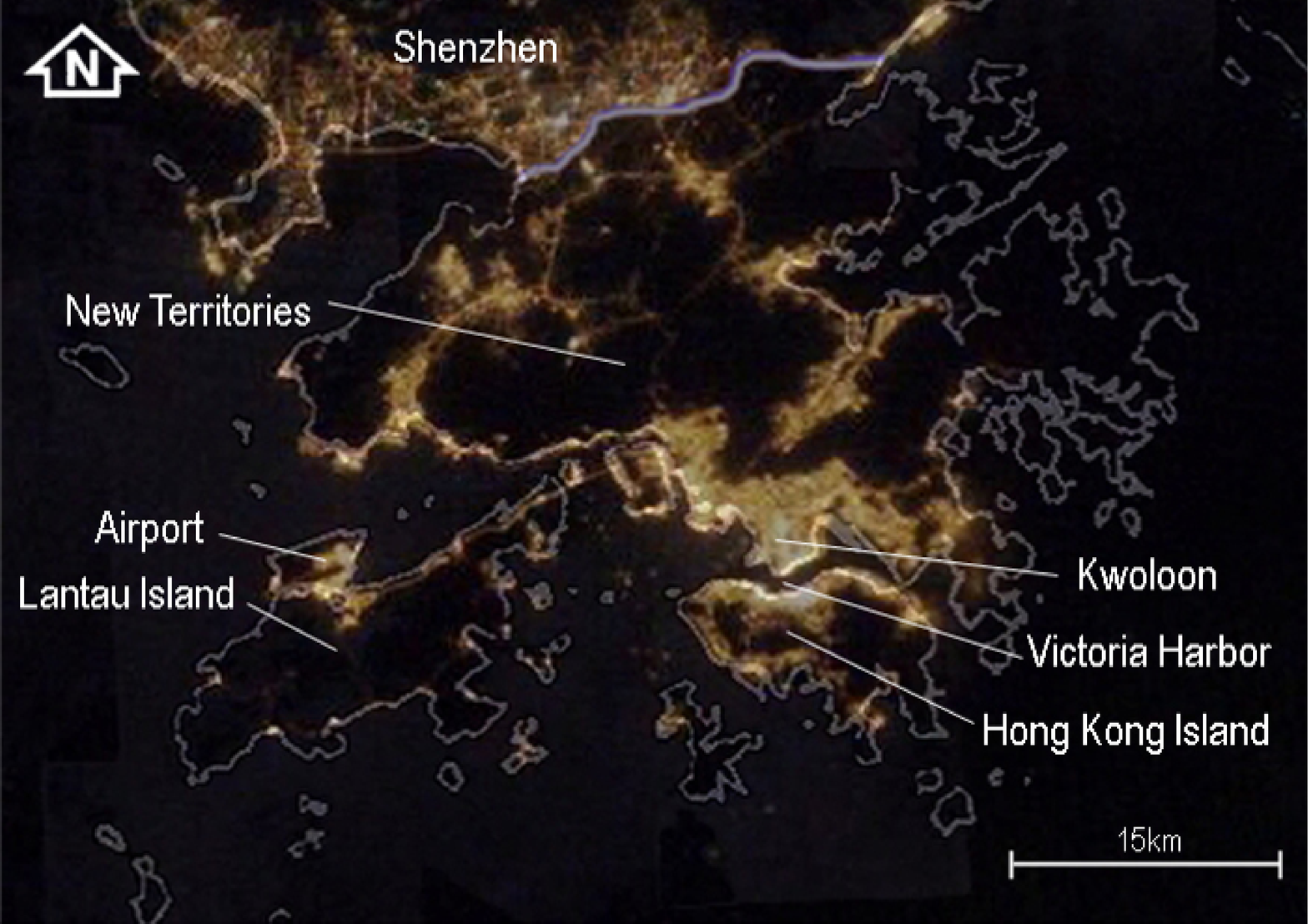}
\caption{The light pollution in Hong Kong is best illustrated by this night-time picture which was taken by astronauts on the International Space Station (ISS) flying 350-400 km above the Pearl River Delta region during 2007-2008. (Credit: NASA)} 
\label{fig:ISS-HK}
\end{figure}

Night-sky brightness (NSB) can be used as an environmental assessment indicator to characterize the relative intensity of light pollution. As the number of stars visible by unaided eyes depends on the level of light pollution, star-count or estimation of the visual limited magnitude, i.e., the magnitude of the dimmest star observable, can be used to assess the night-sky condition of that location \citep{dordrecht:1973}. Programs such as \textit{GLOBE at Night}\footnote{for more information: \url{http://www.globe.gov/GaN/}} ask volunteering participants to report the limiting magnitude of stars in various constellations in the sky. On the other hand, professional astronomers use astronomical telescopes equipped with CCD cameras to capture night-sky images. Using the standard technique of astronomical photometry, the brightness levels of the star-free region(s) on CCD images can be used to determine accurately the NSB from an observatory site \citep{taylor:2004,krisciunas:2007,sanchez:2007,patat:2008,IAO:2008,marco:2009}. Another approach to estimate the level of light pollution is by remote sensing techniques using night-time satellite images covering a wide area (Figure~\ref{fig:ISS-HK}). \citet{chalkias:2006} modeled the light pollution distribution in suburban areas nears Athens, Greece by extracting NSB information from satellite data obtained from the United States Defense Meteorological Satellite Program.

This paper presents a survey of the condition of light pollution in Hong Kong in which accurate NSB data can be obtained over a wide area over a long period of time. Hong Kong is a dense metropolitan city famous for its spectacular night lights. It is very common there to find poorly designed outdoor lightings that are not properly shielded so that a majority of light is directed upwards towards the sky, and not to the area of interest on the ground. With a highly mixed land utilization within a small area, along with a complex landscape and robust human activities, the night-sky condition of Hong Kong makes for an interesting case study for the effect of human activities on the quality of night-sky. The large scale survey in terms of area of coverage had been conducted and it was able to study the long term time variation of the NSB by monitoring it regularly. A portable device called the Sky Quality Meter (SQM)\footnote{for additional information: \url{http://www.unihedron.com/projects/darksky/}} was adopted for accurate measurements of the NSB in the survey. A team of $\sim200$ volunteering participants, made up mostly from young students in high schools, contributed to the survey, making it an educational citizen science project. The method of the survey and data collection will be described in Section~\ref{sec:method}. Results of the survey are presented in Section~\ref{sec:result_analysis}, together with an analysis on the short-term and long-term time variations of the NSB, and how the measured NSB correlates with geographical distributions, population densities, land utilizations, and air pollutant concentrations. Discussions are presented in Section~\ref{sec:discussion} while conclusions from this study are presented Section~\ref{sec:conclusion}.

\section{Method, observations and sources of data}
\label{sec:method}
\subsection{Outline of the Survey} 

The participation of a large number of people is a key element for a comprehensive survey of the light pollution over a vast metropolis such as Hong Kong. Drafting a large group of volunteering participants enabled the survey to vastly expand the geographical coverage to many locations. In order to obtain accurate data from the volunteering team of observers, many of whom did not have sufficient training to carry out complicated scientific experiments, the Sky Quality Meter (SQM), a small and easy-to-use tool, was decided for making the night-sky brightness (NSB) measurements. For a 15-month period from 2008 March to 2009 May, a total of 171 volunteers, made up of secondary school students, campsite workers, and amateur astronomers, participated in the survey. These volunteers contributed 1,957 sets of NSB data from 199 distinct observation sites all around Hong Kong.

In addition to conducting the first light pollution survey, another goal of this project is to introduce the participants and the public about the problem of light pollution and all its damaging consequences. A large and diverse group of volunteers was invited to participate in this survey so that they could all experience first hand the light pollution situation. For the secondary school students, this also provided an opportunity to get a taste of scientific research. In addition, a website dedicated to this survey was created\footnote{\url{http://nightsky.physics.hku.hk/}} which acted both as an educational resource on topics related to light pollution and as an interface for participants of the survey to submit their NSB measurements. This survey was therefore designed and run as a citizen science project. 

\subsection{Sky Quality Meter (SQM)}\label{subsec:sqm}

\begin{figure}
\includegraphics[width=8cm]{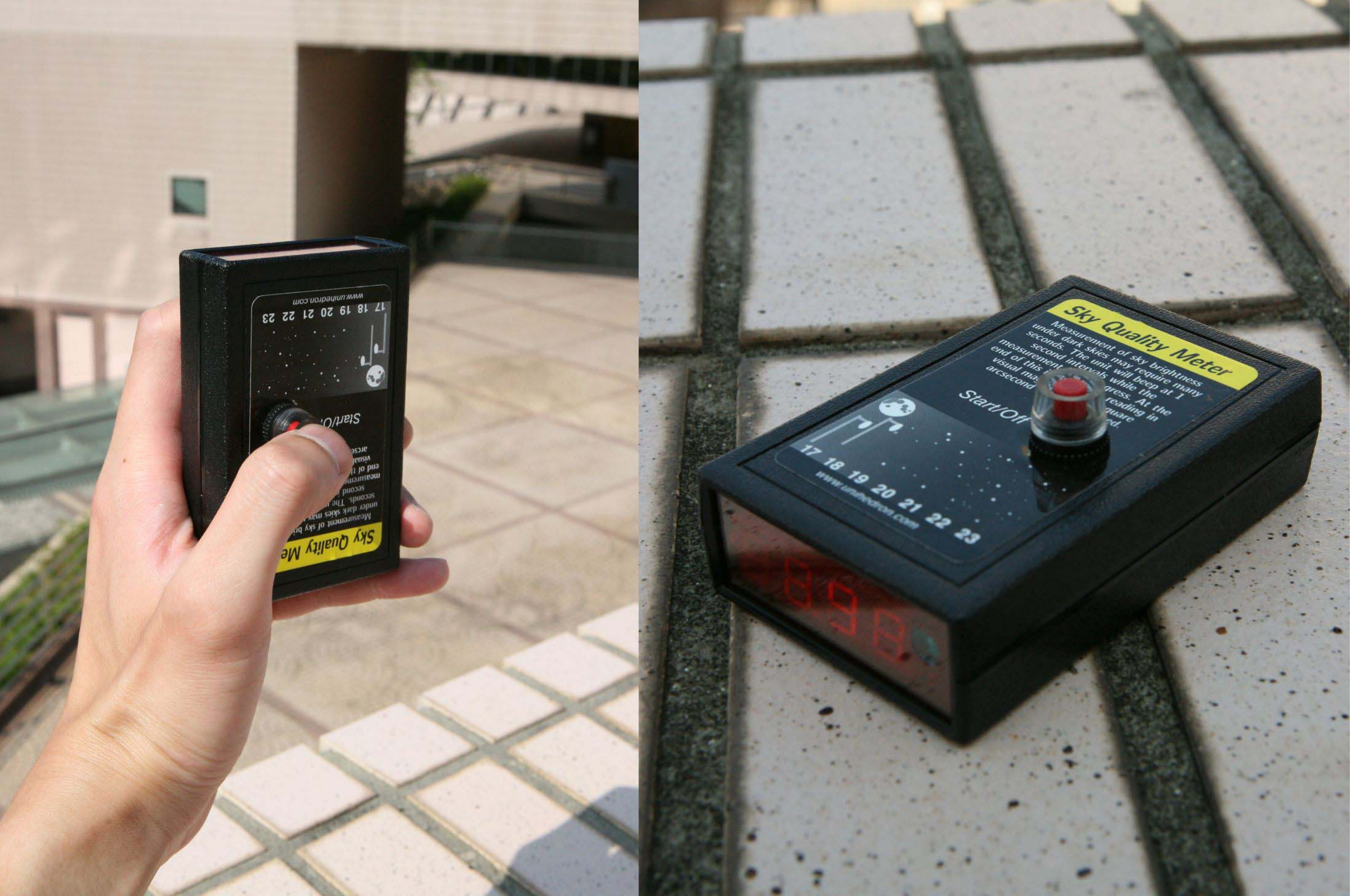}
\caption{The Sky Quality Meter: the light sensor is located on the side of the near-infrared filter next to the display.}
\label{fig:sqm}
\end{figure}

The SQM (Figure~\ref{fig:sqm}) is a device that can instantaneously measure the brightness of the night-sky in units of \textit{magitude per arc second square} (mag arcsec$^{-2}$), the international unit for measuring sky brightness\footnote{\textit{Magnitude} (mag) is a logarithm-scale unit to measure the brightness of astronomical objects. A difference of 1 mag refers to an observed light flux ratio of $10^{-0.4}=2.512$. \textit{Arc second} (arcsec or ") is the unit of length on the celestial sphere. 1" = 1/3,600 degree. Suppose the measured NSB at site A is 20.0 mag arcsec$^{-2}$, then the brightness of the sky is equivalent to a celestial object of 20 mag filling up a patch of sky of area 1 arcsec $\times$ 1 arcsec. Suppose the measured NSB at site B is 19.0 mag arcsec$^{-2}$, then the sky at site B is 2.512 times brighter than that at site A.}. The unit of mag~arcsec$^{-2}$ can be converted to the linear illuminance units \textit{nanoLamberts} (nL)\footnote{1 $\rm nL= 3.18 \times 10^{-10}~cd~cm^{-2}$ \citep{ida:units}} common used by lighting engineers through the relation $B$ = 34.08~exp(20.7233-0.92104$V$), where $B$ is the brightness measured in nL units while $V$ is the night-sky brightness in mag~arcsec$^{-2}$ \citep{garstang:1986}. 

The SQM was selected for this project because of its portability (roughly about that of a deck of playing cards operating on a 9V battery) and its ease-of-use. 
The light sensor of the SQM, the TAOS TSL237 High-Sensivity Light-to-Frequancy Converter\footnote{for additional information: \url{http://www.taosinc.com/}}, is covered by a near-infrared blocking Hoya CM-500 filter\footnote{for additional information: \url{http://www.hoyaopticalfilters.co.uk/}} so that the combined filter-sensor system would have similar sensitivity to that of a human eye. To properly use the SQM to measure the NSB, an observer points the sensor-side of the SQM directly to the zenith and activates the light sensor. Readings of the NSB and the ambient temperature will be displayed in about one second. The SQMs are sensitive only to a narrow cone of the sky directly in front of its sensor. The Half Width Half Maximum (HWHM)\footnote{Half Width Half Maximum (HWHM) is the angular size at which the sensitivity falls by half.} of the sensitive light cone for SQM is 42$^{\circ}$ \citep{sqmreport} while that of SQM-L, the lensed version of the SQM, is 10$^{\circ}$ \citep{sqml_web}.

The narrow angular sensitivities of the SQMs reduces the chance for the NSB measurements to be affected by direct illumination from surrounding lightings. The SQMs were calibrated for their absolute sensitivities by the manufacturer.  During the year-long survey, long-term stability and reliability of measurements were crucial. Therefore, SQMs from the observers were collected roughly every 3 to 4 months to check for detector stability after every phase. The angular acceptance of all SQMs were checked before and after the survey. No significant drop in SQM performance was noticed over time. In addition, all units were shipped to the manufacturer for recalibration after the survey was completed. All SQM units were found to be stable in their performances up to a measured brightness of $\pm$0.2 mag arcsec$^{-2}$ (except one of which the infrared blocking filter was lost, and results taken by this meter had to be rescaled), and included in the uncertainty of all NSB measurements of this survey. 

A total of 39 SQM units and 4 SQM-L units were used in the survey, with SQM-L contributing only a small fraction ($\sim$4\%) data. NSB measurements from the two kinds of meters were analyzed together and were not distinguished. This is possible because the two meters are equipped with the identical light sensor and filter systems. Moreover, the  difference in measurements between them was small under the manufacturer calibration condition and within our observations. Therefore the term "SQM(s)" would hereafter be used to represent both SQM and SQM-L. 



\subsection{Observation time} \label{subsec:obstime}
\begin{figure}
\includegraphics[width=8.5cm]{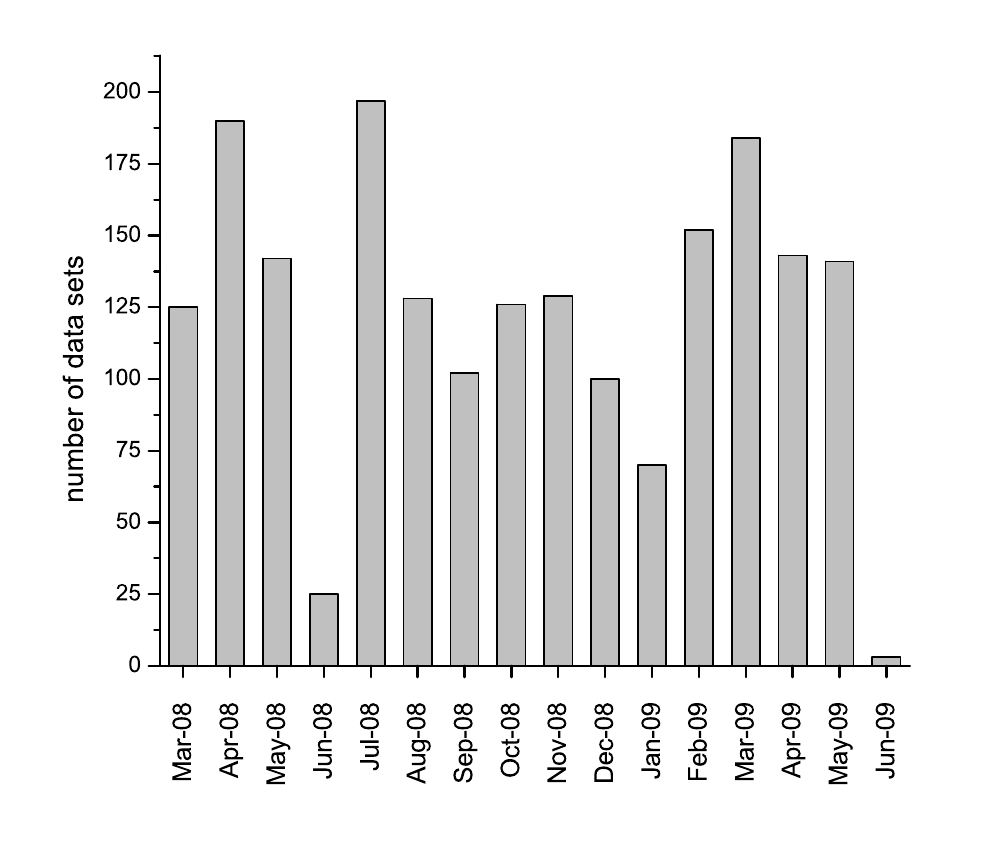}
\caption{Amount of data sets taken versus month. }
\label{fig:data_vs_month}
\end{figure}

The survey started on 2008 March 15 and completed on 2009 May 31. The 15-month duration of the survey was divided into four phases, each of which lasted for roughly 3 to 4 months.
Breaking down the long survey period into phases not only gave more interested parties the opportunity to participate in the survey, but also allowed the SQMs to be checked and calibrated regularly for accuracies. The amount of data sets received in each month of the survey is shown in Figure~\ref{fig:data_vs_month}. Apart from the few months (2008 June and 2009 January) when the SQMs were collected for calibration at the end of each phase of the survey and thus resulting in lower data report rates, consistently over 100 data sets were received each month throughout the duration of the survey.


In order to regularly monitor the full picture of NSB across Hong Kong, while not overworking the volunteering workforce, participants were asked to take NSB readings at a predefined \textit{Survey Time}, which was defined to be 9:30pm and 11:30pm ($\pm$ 5 minutes) on the 5th, 10th, 15th, 20th, 25th, and (if available) 30th day of the month in local Hong Kong Time (HKT, same throughout this article). On the day of the year with the longest daylight (summer solstice) in Hong Kong, the sunset time is around 7:10pm while the evening twilight ends at around 8:35pm. Therefore all measurements in this survey will not be affected by evening twilight. Two observations taken on the same night allowed us to investigate the intranight variation of NSB and the effects of artificial lightings on NSB as many public and private outdoor lights were turned off in late evenings. For safety reasons, participants were advised not to take data under bad weather conditions, for examples, when raining, or typhoon or lightning warning signals were issued. On the other hand, data collected in the survey cover a wide range of natural astronomical (such as Moon phase, Moon/planet/star positions) and meteorological (such as cloudiness and humidity) conditions. Out of the 1,957 data sets collected, 1,023 (52\%) were indeed taken at the specified \textit{Survey Time}. Within that data subset, the number of data sets taken at 9:30pm is slightly higher (57\%) than those taken at 11:30pm.  

\subsection{Observation sites} \label{subsec:obsloc}
One of the strengths of the survey was the deployment of many observers so that NSB readings could be taken simultaneously from many locations across the city. All observers were instructed to take data from at least one location of their own choosing based on the following guidelines: 
\begin{itemize}
	\item Select an outdoor site that is easily accessible, safe, open, and has a wide field of view, such as building rooftops, parks, or playgrounds;
	\item Avoid a site with excessive street lamps or artificial lightings from outdoor buildings or light-emitting signboards nearby so that readings from the SQMs would be minimally affected and the data taken would reflect the true brightness of the sky;	
	\item If there are artificial lightings nearby, measurements should be taken at least 10m from any individual light source.
\end{itemize}
In the majority of cases, these locations are either close to their home (e.g., the secondary school students), their work place (e.g., the campsite workers), or dark sites (e.g., the amateur astronomers or star grazers). In a crowded city like Hong Kong, it is difficult to entirely avoid lightings from street lamps, neon signboards, and buildings in urban or even rural areas. A small percentage of observers might have collected data at inappropriate sites, leading to abnormally bright night-sky observed (NSB < 12 mag arcsec$^{-2}$). However, for the completeness of the study, all the survey results had been included in the large scale analyses as discussed in Section~\ref{sec:result_analysis}. 

\begin{figure}
\includegraphics[width=11.5cm]{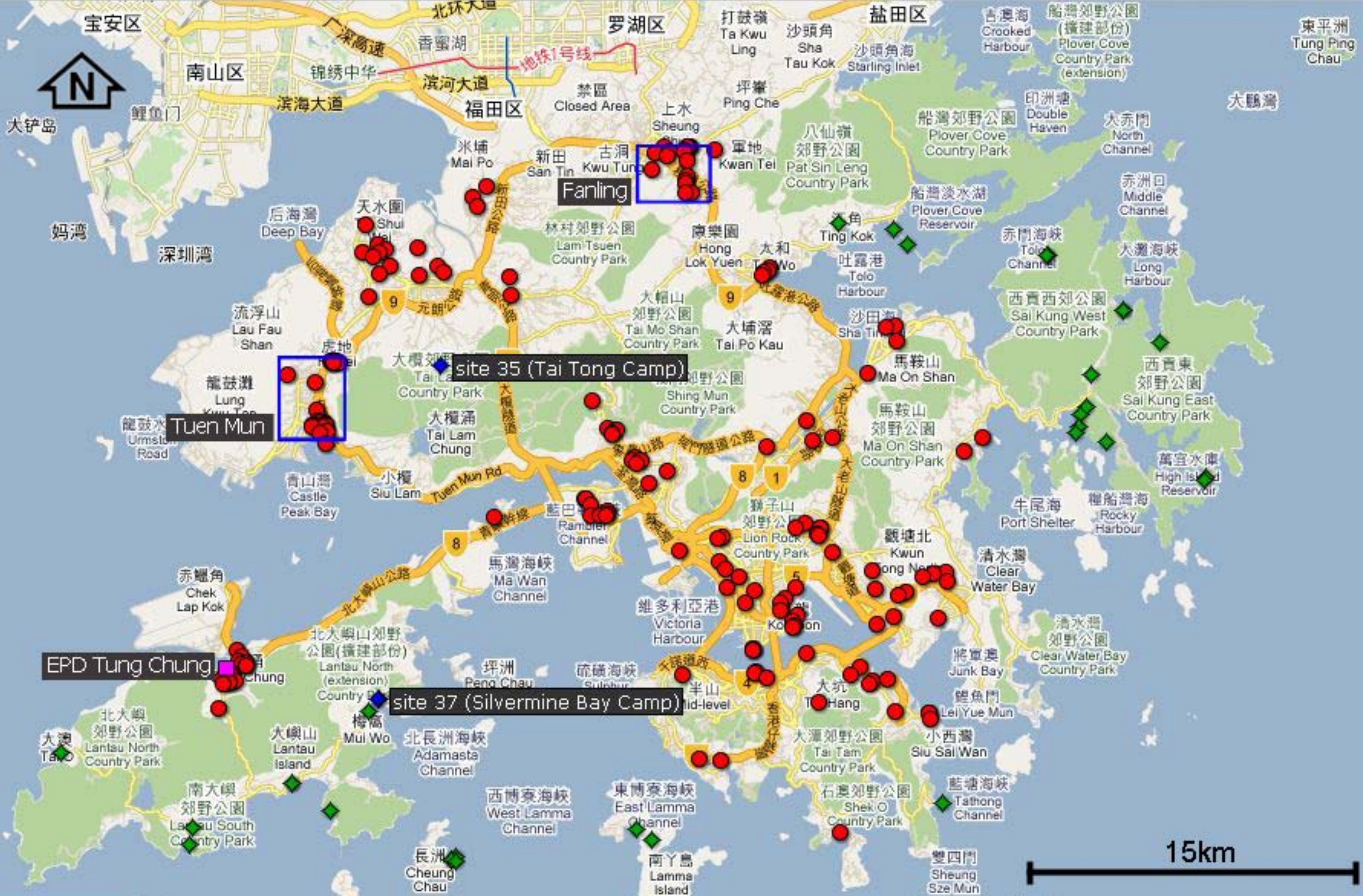}
\caption{Geographic distribution of the observation sites. The urban (red circles) and rural (green diamonds) sites are represented with different symbols. Two blue square frames are the suburban areas (Tuen Mun and Fanling) where the relation between population density and NSB was investigated. (see Section~\ref{subsec:population} for details). The two rural observing stations from which long-term data are studied, site 35 (Tai Tong Camp in Yuen Long) and site 37 (Silvermine Bay Camp in Mui Woo) are also shown (blue diamonds, see Section~\ref{subsec:samesite} for details). Location of the air quality monitoring Tung Chung station of the Environmental Protection Department (EPD) is also marked as pink square (see Section~\ref{sec:air_VS_NSB} for details).}
\label{fig:sitedistrib}
\end{figure}

There were in total 199 individual sites registered by the participants, covering all the 18 Administrative Districts of Hong Kong. The geographical distribution of all the observation sites is shown in Figure~\ref{fig:sitedistrib}. These sites covered a wide range of land utilizations, including GIC (government, institution and community) facilities (some participants were recruited from civic organizations), residential areas, commercial districts, rural settlements, and country parks. All sites were roughly classified into either  \textit{urban} or \textit{rural} categories. Criteria considered for the classification include land utilization and population density of the surrounding areas, which could be estimated from inspections of the neighborhood environments with free online services such as \textit{Google Earth}\footnote{for additional information: \url{http://earth.google.com/}} or \textit{Google Maps}\footnote{for additional information: \url{http://maps.google.com/}}, and the site photos uploaded by the observers. Of the 199 sites, 171 of them (86\% of total) are classified to be urban (represented by red circles in Figure~\ref{fig:sitedistrib}), while the remaining 28 sites (14\% of total), represented by green diamonds in Figure~\ref{fig:sitedistrib}), are classified to be rural. Of the 1,957 data sets of NSB received, 42\% of the data sets were taken at rural sites while the remainings were from urban sites. \\

\subsection{Data taking and reporting}
In order to ensure data of sufficient reliability and accuracy, before the start of each phase, trainings were provided to all participants on the proper procedures to use the SQMs to take NSB data. Detailed instructions\footnote{online copy: \\\url{http://nightsky.physics.hku.hk/web_howto_SQM_eng.pdf}} for data taking and data reporting were also distributed during trainings. For each data set, each observer was requested to take five consecutive readings of the NSB with the SQM so that short-term electronic glitches of the SQMs can be avoided. In addition to the NSB readings from the SQM, each observer was also requested to record the date, time, location, and temperature (read from the SQM) of the observation. Moreover, they were asked to roughly estimate the cloud coverage and the haze condition during the observing time. Other meteorological elements such as humidity had not been recorded.\\

To maintain speedy and accurate reportings of results, an online data reporting system accessible through the project webpage was created. Each participant was given a username and password to report the data of the survey. Participants were also required to provide the latitude and longitude of each site to the survey's database so that all observation locations would be uniquely identifiable using online services such as \textit{Google Maps}. Finally, observers were recommended to upload day-time and night-time photographs of the nearby environment of each site so that the suitability of the site for taking NSB data could be evaluated. \\

\section{Results and analysis}
\label{sec:result_analysis}
\subsection{Overall results of Hong Kong} \label{subsec:overall}

\begin{figure}
\includegraphics[width=8cm]{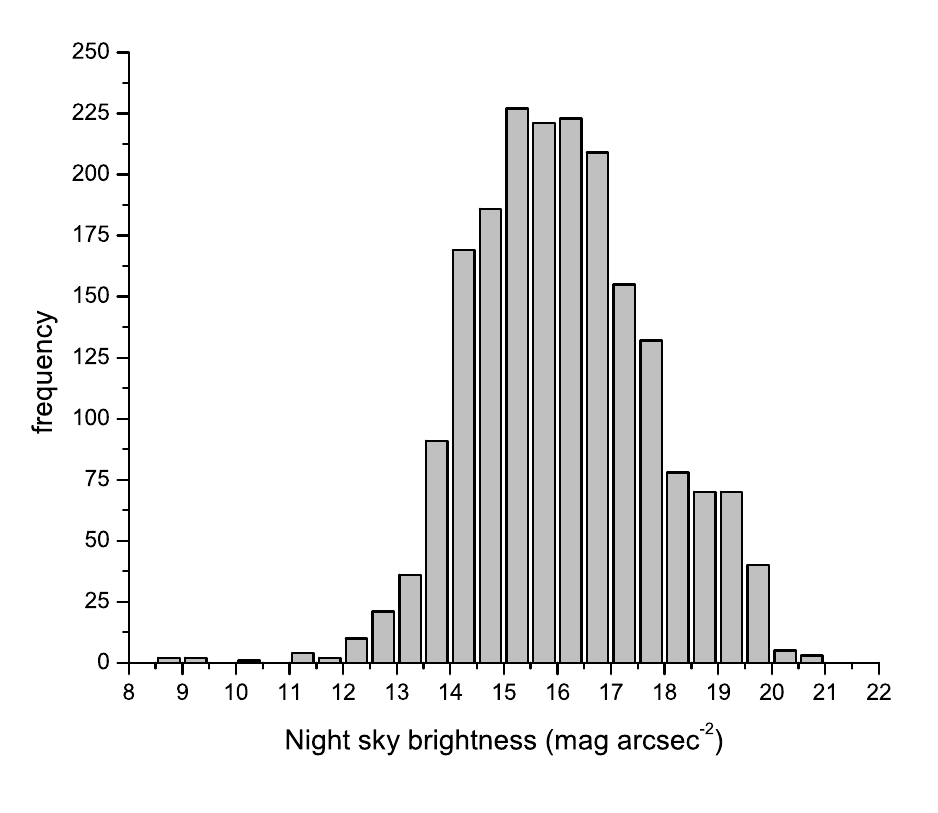}
\caption{Histogram of the brightness of the night-sky collected in the survey.}
\label{fig:histogram_NSB}
\end{figure}

With contributions from 171 volunteer participants, the survey received a total of 1,957 sets of NSB data collected at 199 distinct observation sites during the entire duration from 2008 March 15 to 2009 June 1. The histogram of all the NSB readings collected are shown in Figure~\ref{fig:histogram_NSB}. The distribution peaks at around 15.5~$-$ 16.5 mag~arcsec$^{-2}$. The darkest night-skies were detected in Shui Hau in the southern Lantau Island and at the East Dam of the High Island Reservoir in eastern Sai Kung with a measured NSB value of 20.8 mag arcsec$^{-2}$. Both these locations are rural areas adjacent to country parks away from population settlements. On the other hand, any measured NSB readings smaller than 12 mag arcsec$^{-2}$ are probably be due to either incorrect data input (human error during data reporting on the web), incorrect usage of SQM (such as not orienting the light sensor along the zenith), or improper site selection (that is, measuring at sites with external lightings directly entering into the detection cone of the SQM light sensor). These data make up only a tiny portion (0.61\%) of the total survey database. 

The overall average NSB reading obtained in this survey is 16.1 mag~arcsec$^{-2}$. However, this number should be treated with caution because the data had been collected at vastly inhomogeneous conditions. Effects of factors such as time of observations, locations of study, and air pollution factors will be studied and presented in the following sections. On the other hand, this figure does present a characteristic level of night-sky brightness in Hong Kong as the survey covered a wide variety of locations and a wide range of time frames under diverse natural and artificial conditions. The current level is significantly brighter than the natural NSB level of 21.6 mag~arcsec$^{-2}$ (a 5.5 mag~arcsec$^{-2}$ difference, or 160 times brighter in flux) recommended by the International Astronomical Union (IAU) for conducting astronomical observations \citep{smith:1979}, not to say the darker skies of 21.7~-- 22.0 mag~arcsec$^{-2}$ measured at research observatory locations \citep{marco:2009,patat:2008,sanchez:2007,krisciunas:2007,pilachowski:1989}. On the other hand, the darkest sites in Hong Kong at 20.8 mag arcsec$^{-2}$ remain to be reasonably good sites for public enjoyment on the starry night and for astronomy work despite the heavy urbanization in many parts of the city and it is essential that efforts should be made for these locations to remain so as discussed in Section~\ref{intro}.

\subsection{Geographic variation of night-sky brightness}\label{sec:geovariation}

One key factor affecting the measured NSB is the observing location. This is expected because of the vastly different population densities, land utilizations, outdoor lighting environment and practices in different areas in Hong Kong. 
The NSB distribution around Hong Kong using the entire data set will be overviewed in Section~\ref{subsec:HKmap}. The relation between the observed NSB and population densities will be studied in Section~\ref{subsec:population}.  Snapshot of the NSB distribution taken at a specific date will be presented in Section~\ref{subsec:snapshots} while detailed studies of NSB distribution against land utilizations will be surveyed for the region of Sai Kung in  Section~\ref{subsec:landuse}.

\subsubsection{The Hong Kong Light Pollution Map} \label{subsec:HKmap}

\begin{figure}
\includegraphics[width=11.5cm]{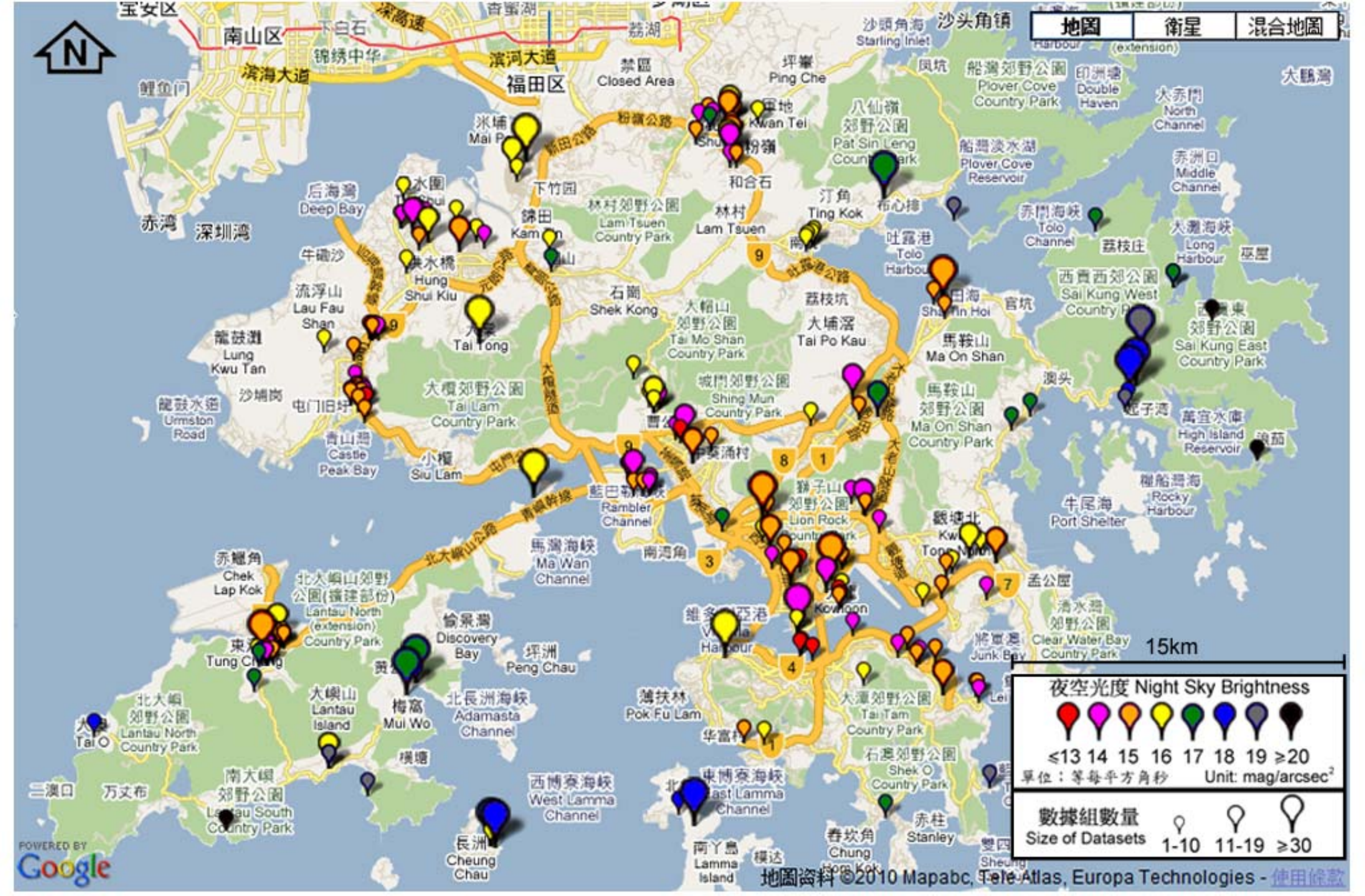}
\caption{The Hong Kong Light Pollution Map using \textit{Google Maps}. Icon colors represent the averaged NSB measured for the sites. Red and purple icons (small NSB values) represent brighter sky while gray and black icons (large NSB values) represent darker sky. The size of the icon represents the actual number of data sets (one data set contains five consecutive NSB readings) collected for that site. The tip of icons marked the exact location of sites.}
\label{fig:googlemap}
\end{figure}

The geographical dependence of the brightness of night-sky in Hong Kong was presented by constructing the Hong Kong light pollution map as shown in Figure~\ref{fig:googlemap}. For each observing site, the average value of all observations taken from that location was computed and marked on a map that was generated with the free online service \textit{Google Maps}. The average NSB at each location is represented with different color symbols to show the variation of the observed NSBs, where smaller value represents brighter sky while larger value represents darker sky. As it will be discussed below in Section~\ref{subsec:timevariation}, the NSB at any location varies significantly over time and thus it could be misleading to represent the NSB of any location with its averaged value. 

Moreover, the accuracy and significance of the average NSB shown in Figure~\ref{fig:googlemap} varies significantly for each location, ranging from sites with large number of observations (187 data sets for site 35) to the 38 sites with only one single reported measurement. In Figure~\ref{fig:googlemap}, the size of the icon indicates the number of data set (one data set contains five consecutive NSB readings) collected at each site. An interactive version of the light pollution map is also available online on the survey website, with which visitors can zoom in on their neighborhood of interest and check the measured NSB taken near these locations. 

The study indicates that the quality of night-sky in Hong Kong is strongly associated with the human presence and human activities. High population and thus high lighting densities in dense urban areas cause severe light pollution. 
Comparing the survey locations and the NSB measurements in Figures~\ref{fig:sitedistrib} and \ref{fig:googlemap} indicate that there is a general relation between the night-sky quality with the rural/urban land use classification. Most of the sites with dark night-skies (NSB readings at or above 17 mag~arcsec$^{-2}$) have been classified to be in rural locations. On the other hand, bright night-skies can be found on both sides of the Victoria Harbour, the traditional center of the city with the highest level of urban developments. The hilly landscape of Hong Kong also implies that there could be large difference in the observed night-skies between locations at only a few kilometers apart. 

The brightest (red icons) sites recorded in this survey were two locations Mong Kok in Kowloon and Wan Chai in the northern Hong Kong Island, both yielding averaged NSB readings of 13.2~mag~arcsec$^{-2}$. Comparing this with the darkest averaged night-skies (black icons) at 20.1~mag~arcsec$^{-2}$ (also at Shui Hau and East Dam, where the darkest individual readings were recorded), the difference of 6.9~mag~arcsec$^{-2}$ in NSB measured suggested that the brightest night-sky in Hong Kong can be over 550 times brighter than that at the most pristine dark sky available. On average, the NSB value for all the 171 urban sites at 15.0 mag~arcsec$^{-2}$ suggests that the urban night-sky is on average 100 times polluted by outdoor lightings compared to the darkest night-sky observable in Hong Kong. 


Special care should be taken when interpreting the light pollution map presented in Figure~\ref{fig:googlemap}. As will be discussed below in Section~\ref{subsec:timevariation}, in addition to the observation location, the measured NSB value at a particular site also depends on the date and time of the observation. Therefore the light pollution map presented here would not believe to be an accurate predictor of the NSB of any specific location on a certain night, nor do anyone put high significance on all the detailed distribution of NSB within a small region, particularly for cases where large NSB differences are observed for adjacent sites because data from adjacent sites could have been taken from different nights under different conditions. On the other hand, with the big number of data sets received covering large fractions of land area, it is believed that this light pollution map does provide a general overview of the light pollution situation in Hong Kong and can serve as a starting point for the understanding of the environmental light pollution problem.  


\begin{figure}
\includegraphics[width=11.5cm]{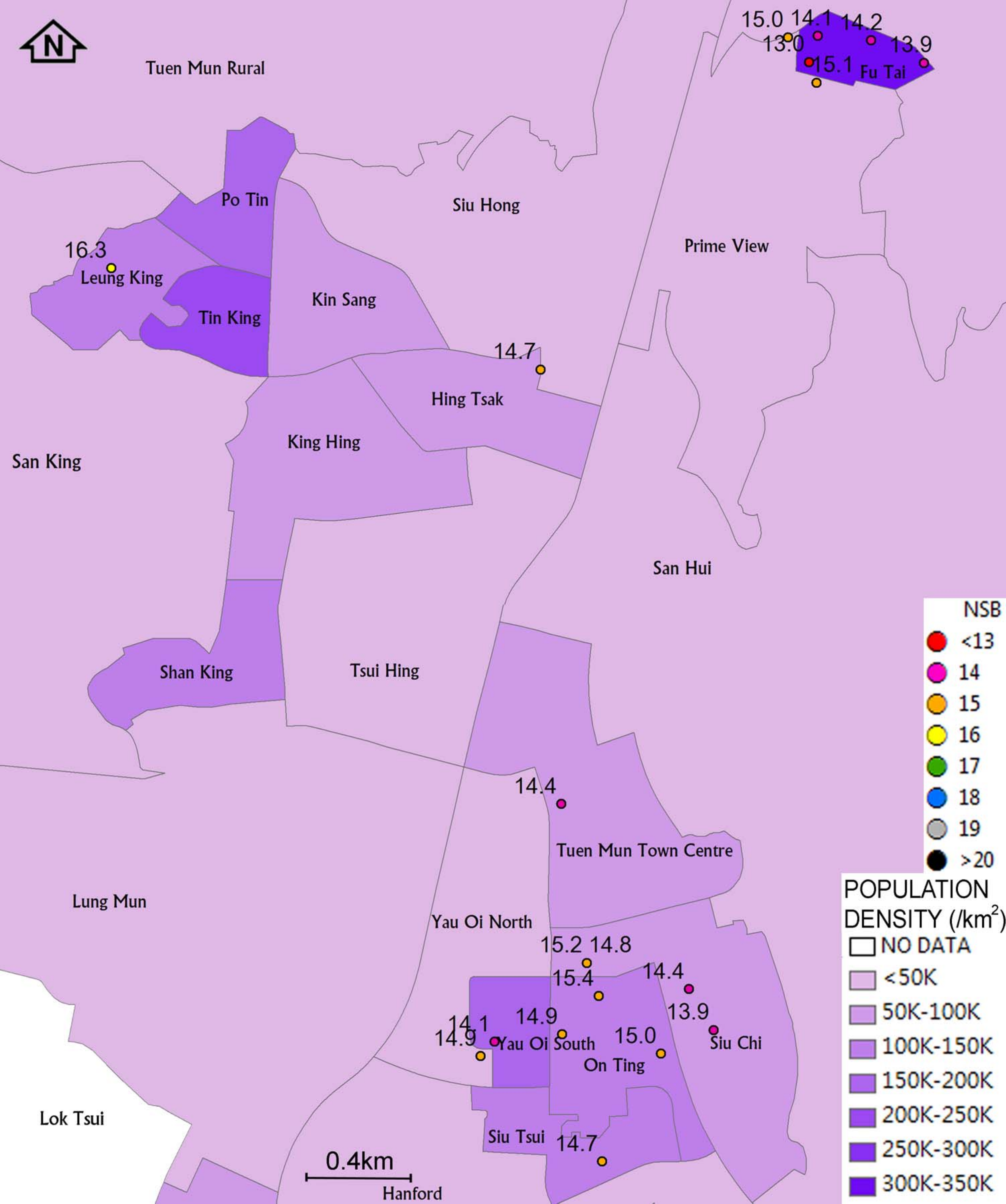}
\caption{Population density and night-sky brightness in Tuen Mun located in western Hong Kong (as marked by the blue square frame in Figure~\ref{fig:sitedistrib}). Locations of the observation sites are marked by circles with the value of the averaged NSB marked. Red and purple icons (small NSB values) represent brighter sky while gray and black icons (large NSB values) represent darker sky. Sites with average NSB readings smaller than or equal to 12.5 mag~arcsec$^{-2}$ were removed. Population density of each DCCA is shown in different shades.} 
\label{fig:TuenMun}
\end{figure}

\begin{figure}
\includegraphics[width=10.5cm]{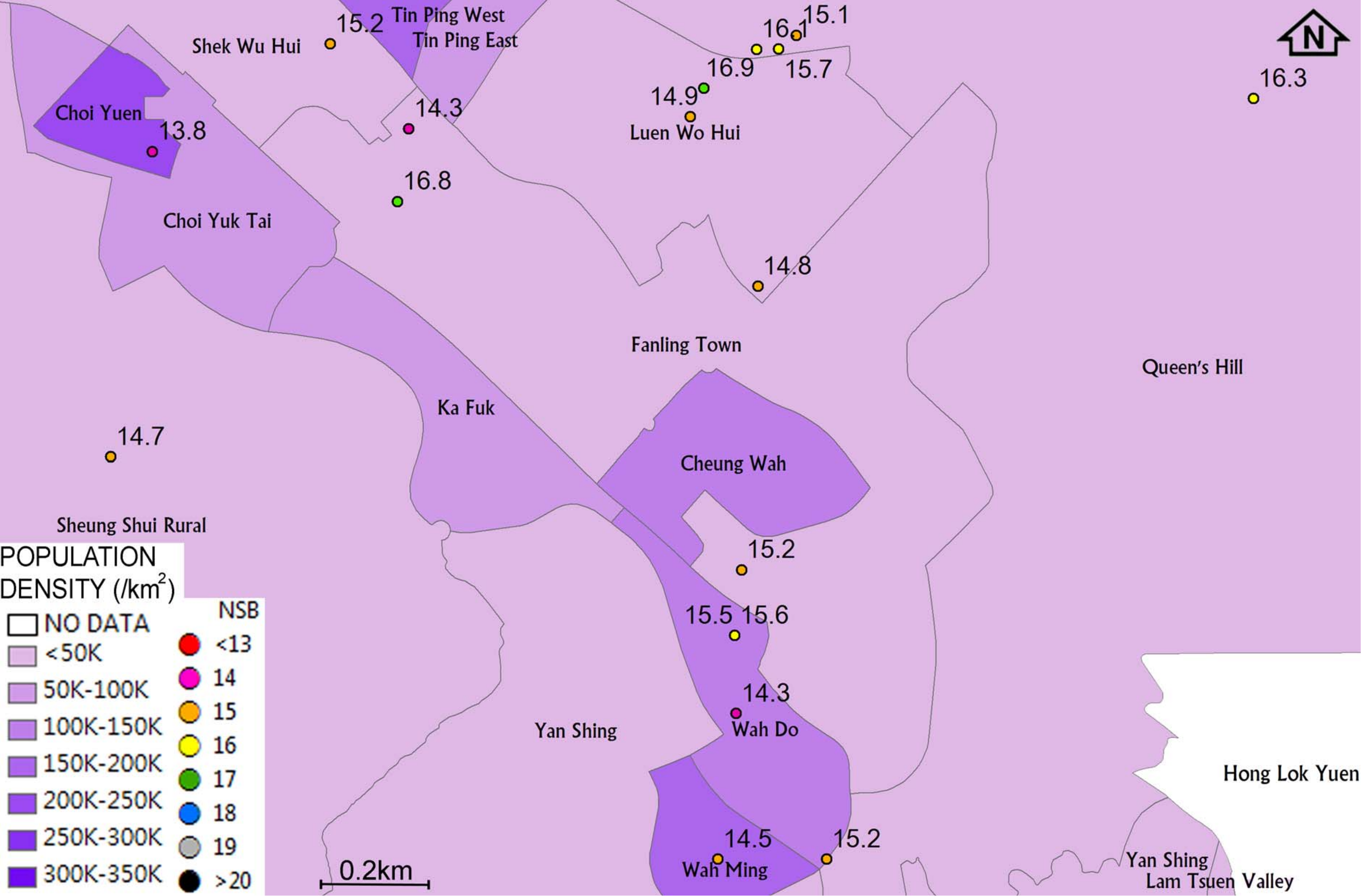}
\caption{Same as Figure~\ref{fig:TuenMun} for Fanling in northern Hong Kong.}
\label{fig:Fanling}
\end{figure}

\subsubsection{Population density and night-sky brightness} \label{subsec:population}

The city of Hong Kong is divided into 400 District Council Constituency Areas (DCCAs) which represent the smallest population units surveyed in the Hong Kong 2006 Population By-census\footnote{for additional information: \url{http://www.bycensus2006.gov.hk/en/index.htm}}. DCCA boundaries had been declared by the Registration and Electoral Office for the 2003 District Council Election\footnote{for additional information: \url{http://www.elections.gov.hk/elections/dc2003}}. The DCCAs were established such that the size of each District will have roughly the same population, leading to large variations in terms of their sizes. As discussed before, the complex landscape of Hong Kong implies that the NSB measured from different locations within the same DCCA can be significantly different. Therefore it was decided not to study the variation of NSB with the population directly throughout the city.  

To study the influence of human activities on the degradation of the dark sky, two suburban population centers, Tuen Mun on the western side and Fanling on the northern side of Hong Kong (locations highlighted by the blue square frames in Figure~\ref{fig:sitedistrib}), were selected instead. These two regions are selected because of the large number of observations in multiple locations taken, and because of their relatively flat landscapes. In Figures~\ref{fig:TuenMun} and~\ref{fig:Fanling}, the population densities of the DCCAs obtained from the 2006 Population By-census were overlaid on the values of the averaged NSB measured in Tuen Mun and Fanling respectively. A weak trend of increasing NSB with population density was noticed. The trend is not stronger here probably because within a densely populated environment like Hong Kong, the NSB at a darker DCCA can be heavily affected by scattered lights from the bright DCCA(s) nearby. This effect is particularly important at the borders between multiple DCCAs. Moreover, as will be discussed below, NSB is a sensitive variable which depends highly on other environmental and atmospheric conditions, and can fluctuate in the order of 1 mag~arcsec$^{-2}$ at the same location. A limit of this survey was that that the volunteers were free to select their observing sites, the survey would not be able to ensure that every population density level had an equal chance of having an observation.


\subsubsection{Snapshot of the night-sky in Hong Kong} \label{subsec:snapshots}

\begin{figure}
\includegraphics[width=10.5cm]{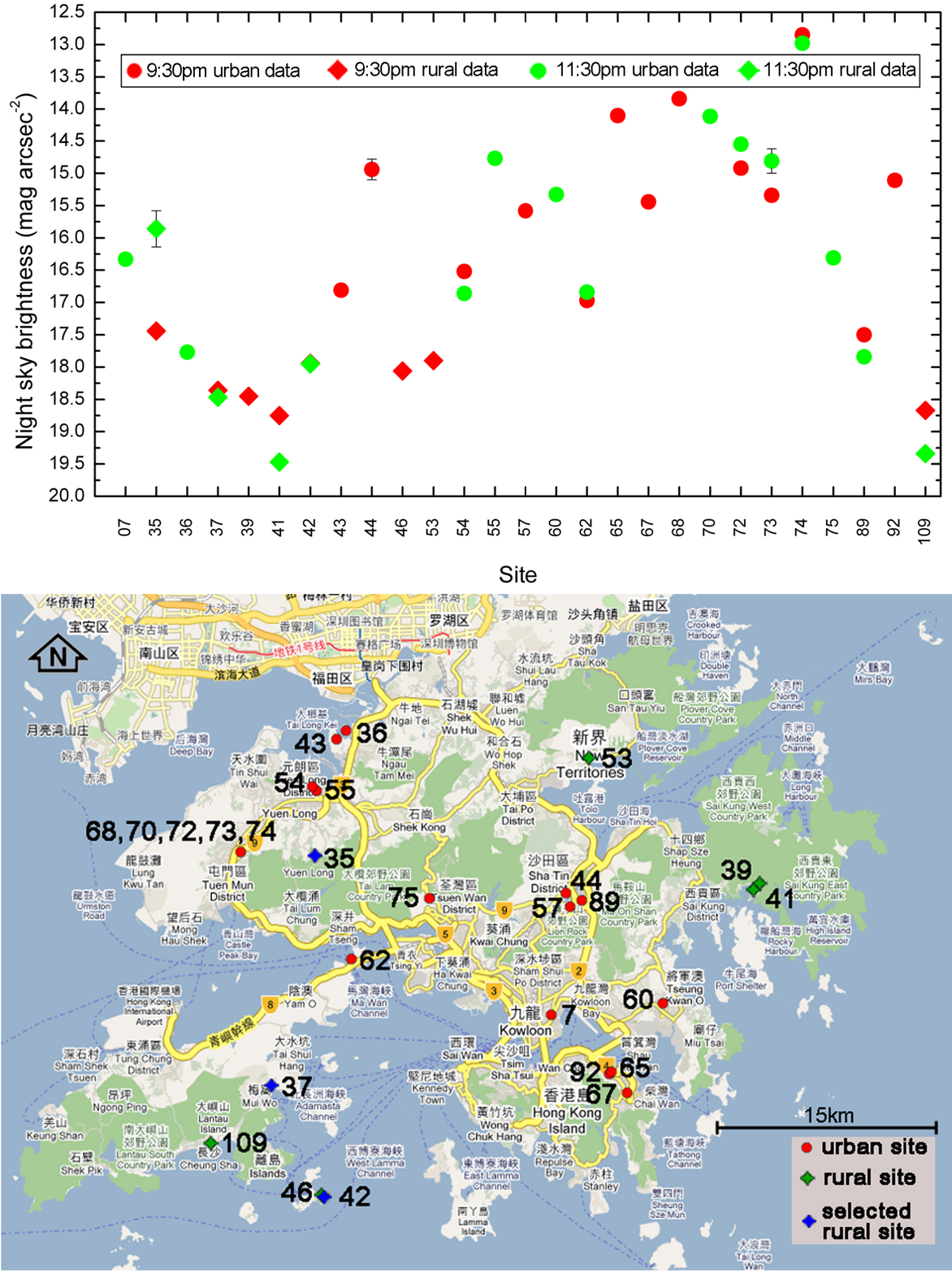}
\caption{Snapshot survey of the night-sky brightness (NSB) in Hong Kong on 2008 April 5. Simultaneous NSB measurements taken at different observing sites are plotted in the upper panel, with the exact locations of the sites shown in the bottom panel. The data from urban and rural sites are represented with circle and diamond symbols respectively, with NSB readings taken at 9:30pm and 11:30pm marked in red and green respectively in the upper panel. Small NSB values represent brighter sky while large NSB values represent darker sky. Error bars shown are the standard deviations of the five SQM readings for every data set. Notice that the same NSB values was measured at site 42 for 9:30pm and 11:30pm.}
\label{fig:sametime_05Apr08}
\end{figure}

In order to study the geographical distribution of the night-sky condition, one need to isolate the effects due to observation locations by setting all other parameters to be identical. To this end, a date with the most data sets taken from the largest number of distinct locations was selected. The measured NSB readings for observations taken on 2008 April 5 are shown in Figure~\ref{fig:sametime_05Apr08}. The Moon set at 5:51pm on that night and thus moonlight, the dominating natural night light contributor, was not a factor in this set of measurements. In the figure, the top plot shows the NSB of that night at the various observing sites, the exact locations of which are shown in the map below. The sites, as described in Section~\ref{subsec:obsloc} above, are classified into urban and rural sites and are represented respectively by circle and diamond symbols. The NSB readings taken at the two survey times, 9:30pm and 11:30pm, are represented with red and green icons respectively. The error bars shown in the plot are the standard deviations of the five SQM readings reported by the observers for every data set. Benefited from the advantage that the observations were taken under similar conditions, a realistic snapshot of the condition of the night-sky in Hong Kong is obtained from Figure~\ref{fig:sametime_05Apr08}, compared to the general Hong Kong light pollution distribution presented in Figure~\ref{fig:googlemap}.


From Figure~\ref{fig:sametime_05Apr08}, the observed NSB vary significantly among different locations in Hong Kong. On 2008 April 5 at 9:30pm, the brightest observed night-sky at site 68, located within a public housing estate in Tuen Mun\footnote{location: 22$^\circ$24'47"N 113$^\circ$59'10"E} at 13.8 mag~arcsec$^{-2}$, was 5.0 mag~arcsec$^{-2}$, or 100 times in light intensity, brighter than the darkest location at site 41, a camping site located inside a country park in Sai Kung\footnote{The Boys' and Girls' Clubs Association of Hong Kong Bradbury Camp, location: 22$^\circ$23'30"N 114$^\circ$19'07"E}, with 18.8~mag~arcsec$^{-2}$. (The result from site 74 was excluded as the measured NSB was much brighter than from the surrounding locations of sites 68, 70, 72, and 73.) Similar large spread is observed for data taken at 11:30pm. The huge difference in NSB measured in different locations in Hong Kong at the same time suggests that the surrounding lighting environments of the observing sites can have significant effects on the quality of the night-sky. 

Another evidence of the human influence of the night-sky is that the brightness of the night-sky at the 19 urban locations (circles in Figure~\ref{fig:sametime_05Apr08}) are almost all brighter than that observed at the 8 rural locations (diamond symbols). This is true for observations taken at both 9:30pm and 11:30pm, with site 35 as the only exception. However, it is suspect that other factors might have affected the NSB of this location on that night because the observed night-sky at 11:30pm was over 1.5 mag~arcsec$^{-2}$ brighter than that at 9:30pm, by far the biggest brightening observed in a single evening among all locations. It is believed that the urban-rural difference strongly indicates that human activities in urban areas are essential on the quality of the night-sky, and the extent of these activities is the key factor in determining the wide ranges of the NSB observed in different locations around Hong Kong at the same time. 

In Figure~\ref{fig:sametime_05Apr08}, there were 11 sites where data were taken at both 9:30pm and 11:30pm. For all the rural sites (except the problematic site 35 described above), the night-sky at 11:30pm was dimmer than that at 9:30pm. It can be explained by the switching off of artificial lightings for human activities in the later evening time. However, there was a mixed relation between the brightness of the 9:30pm and 11:30pm night-sky for the urban sites, with a half actually showing a brighter night-sky at a later time of the evening. It is probably due to the high number of public and private external lightings used in the city and thus the brightness of the night-sky depends on the complex illuminating and usage patterns of these light sources. 

\subsubsection{Land utilization and night-sky brightness} \label{subsec:landuse}

\begin{figure}
\includegraphics[width=11.5cm]{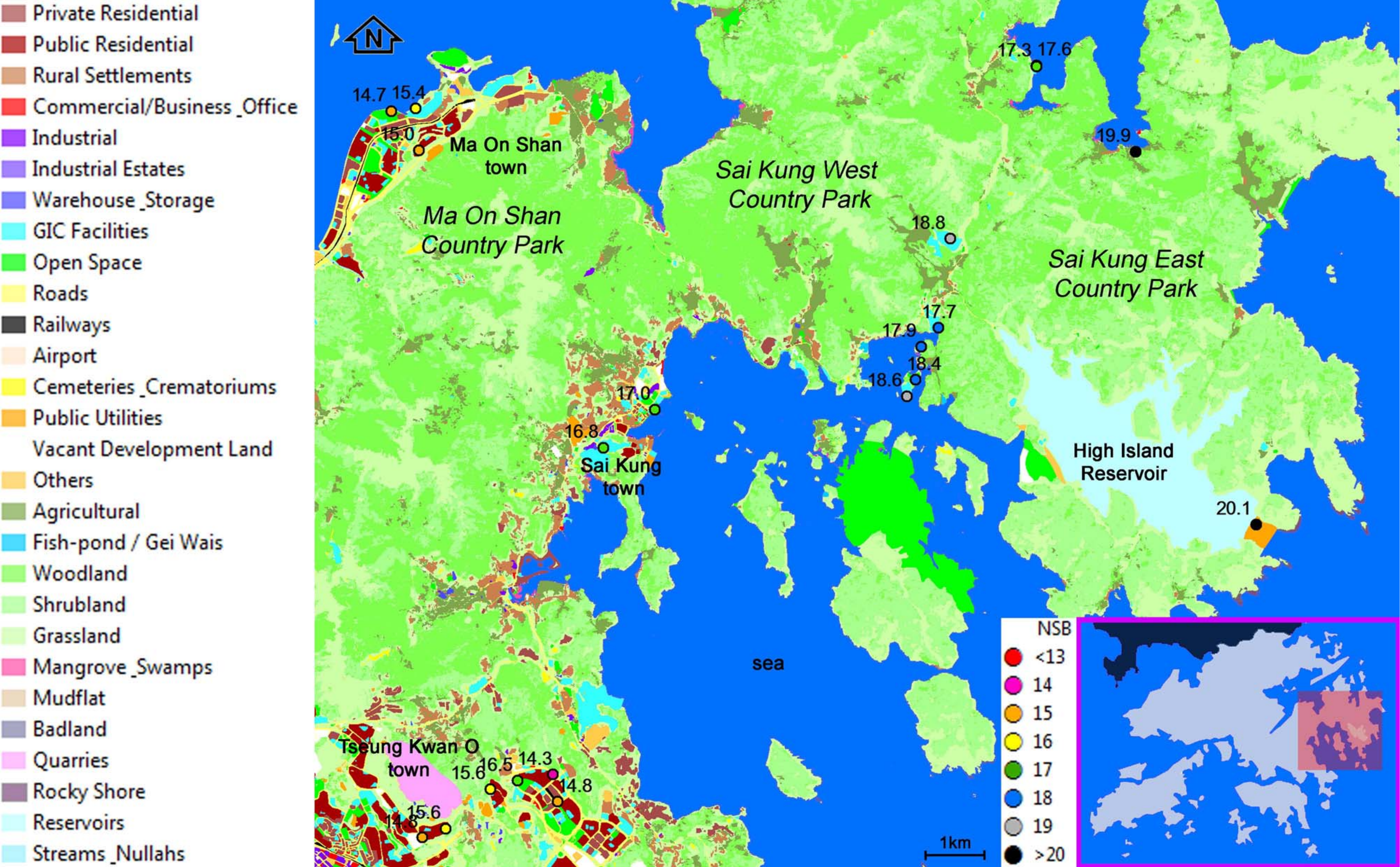}
\caption{The map of land utilization overlaying on the light pollution map in Sai Kung (eastern New Territories). Locations of circles are sites where NSB observed and the neighboring values are the NSB averages. Red and purple icons (small NSB values) represent brighter sky while gray and black icons (large NSB values) represent darker sky. Sites with average NSB readings smaller than or equal to 12.5 mag~arcsec$^{-2}$ were removed. Polygons represents land utilizations. The trends of decreasing NSB (increasing in value) from the west (filled with red colors of residential and commercial regions) to the east (dominated by green woodlands, shrubland, or grasslands) was clear.}
\label{fig:land_use_saikung}
\end{figure}

To study the relation between the quality of the night-sky and the mode of land use, a region in eastern Hong Kong centered on the rural town of Sai Kung (location found in the inserted map of Figure~\ref{fig:land_use_saikung}) was selected for a detailed study. The land usage in Hong Kong is defined by the Planning Department of the government and any piece of land can be classified into any one of 28 land utilization categories. As seen in Figure~\ref{fig:land_use_saikung}, the region being studied is dominated by the woodland/shrubland/grassland of the country park. The population centers include the rural town of Sai Kung near the center and the two suburban towns Ma On Shan and Tseung Kwan O in the northwestern and southwestern corners of the map respectively. Each color region in Figure~\ref{fig:land_use_saikung} represents the officially classified land utilization\footnote{the 2008 land utilization map is provided by the Planning Department (for additional information: \url{http://www.pland.gov.hk/})}. Overlaid in the map are the NSB observation sites (color-coded circles in Figure~\ref{fig:land_use_saikung}) together with the average NSB values measured. 

This study indicates that the quality of night-sky in Hong Kong is strongly associated with human presence and human activities, as predicted by the different land utilizations. This is probably caused by the vastly different outdoor lighting environment and practices in existence in these different regions. In Figure~\ref{fig:land_use_saikung}, there is a gradual trend of darkening of the night-sky as we move from the suburban town on the northwestern and southwestern corners (dominated by the Private and Public Residential land uses), to the rural town in the center (dominated by Rural Settlements land use), and finally to the large patches of greenlands in the country park. In general, the farther the distance from the population centers, the darker the night-sky one can expect. This can be illustrated by the darker night-sky observed on the eastern side of the High Island Reservoir compared to the western side. The study indicates that land use can possibly serve as a indicator of the quality of the night-sky. On the other hand the influence of night lightings in one land utilization region to the adjacent areas had not been investigated in detail. A more detailed study of the quality of the night-sky, in particular in regions between vastly different land uses, may provide hints on the usefulness of this idea.
 
\subsection{Time variation of night-sky brightness}\label{subsec:timevariation}

\subsubsection{Intranight variation of night-sky brightness at urban and rural sites}  \label{subsec:timesurvey}
The variation of the night-sky brightness within the same evening was investigated by comparing the NSB data taken at different times of the night. The NSB of Hong Kong at 9:30pm averaged from 766 sets of data (include urban and rural sites) is found to be 16.1 mag~arcsec$^{-2}$, while the average at 11:30pm from 605 data sets is 16.7 mag~arcsec$^{-2}$, implying a consistently dimmer night-sky by 0.6 mag~arcsec$^{-2}$ at a later time of the evening.  On the other hand, as described in the snapshot survey in Section~\ref{subsec:snapshots}, very different intranight time variations for urban and rural locations were observed. Using the data collected throughout the survey, the respective difference for urban-only and rural-only sites are 0.1 and 1.1~mag~arcsec$^{-2}$ respectively, corresponding to the averaged 9:30pm night-sky to be 10\% and 280\% brighter than 11:30pm in terms of light intensity. The different observed behavior is probably due to the different lighting patterns of the external public and privite light sources, with a majority of the lightings remained lit up to serve the large number of people still engaging in outdoor activities at late night. On the other hand, it is encouraging to learn that the larger contrast of observed NSB readings between 9:30pm and 11:30pm in the rural areas can probably be attributed to external lightings being turned off in responsible manners in later evenings.

\subsubsection{Long-term variation of night-sky brightness at selected sites} \label{subsec:samesite}

Two sites were selected to investigate whether any long-term trend of NSB can be detected. The two sites are site 35 in northwestern Hong Kong near the suburban town of Yuen Long\footnote{Po Leung Kuk Jockey Club Tai Tong Holiday Camp, location: 22$^\circ$24'42"N 114$^\circ$01'52"E, data taken at an outdoor car park and lights from lamps were avoided} and site 37 on the eastern shore of the mostly rural (except the northern shore where the airport is located) Lantau Island\footnote{Hong Kong Playground Association Silvermine Bay Outdoor Recreation Camp in Mui Woo, location: 22$^\circ$16'23"N 114$^\circ$00'12"E, data taken at an outdoor playground with some nearby outdoor lighting turned off at everyday by 11:30pm}. Their geographic locations are indicated by blue diamond symbols in Figure~\ref{fig:sitedistrib}. These two sites are chosen because they have the highest number of pairs of NSB data taken at both 9:30pm and 11:30pm. The volunteering observers also confirm that the lighting environments in the immediate vacinity of the observing sites, including the number of external lightings, lighting distributions, turn-on and turn-off time of lightings, and the lighting intensity did not significantly vary throughout the period of this survey. 


\begin{figure}
\includegraphics[width=11.5cm]{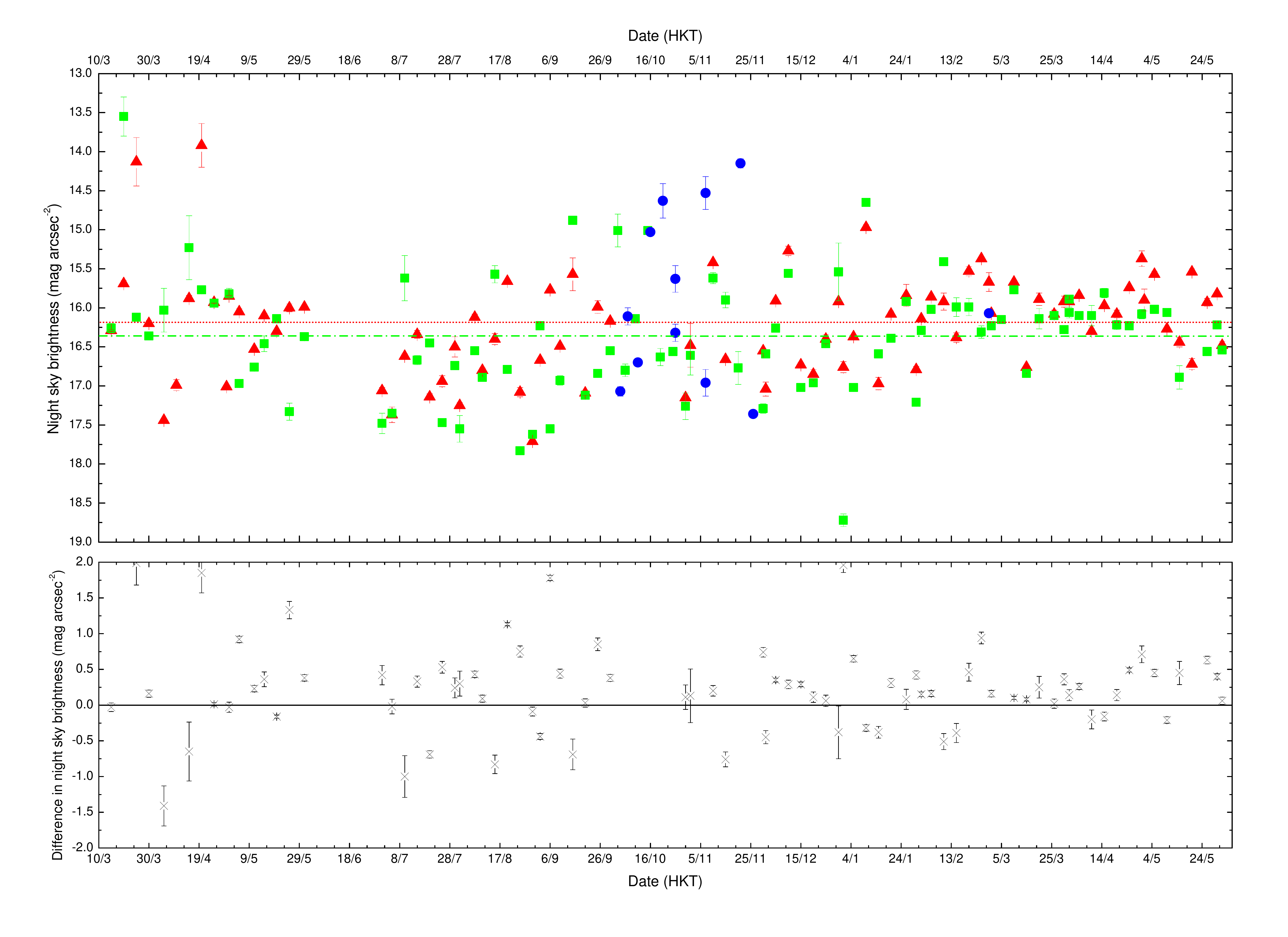}
\caption{\textit{Upper panel}: Time variation of night-sky brightness (NSB) at site 35 from 2008 March to 2009 May. The measured NSB at 9:30pm (red triangles), 11:30pm (green squares), and other times (blue circles) are shown separately. Small NSB values represent brighter sky while large NSB values represent darker sky. Error bars shown are the standard deviations of the five SQM readings for every data set. The survey-average NSB level at this site measured at 9:30pm (red dotted line) and 11:30pm (green dashed line) are also plotted for comparison. \textit{Lower panel}: The intranight NSB difference between 11:30pm and 9:30pm: positive value indicates that the night-sky was darker at 11:30pm and vice versa.} \label{fig:site35_delta_m}
\end{figure}

\begin{figure}
\includegraphics[width=11.5cm]{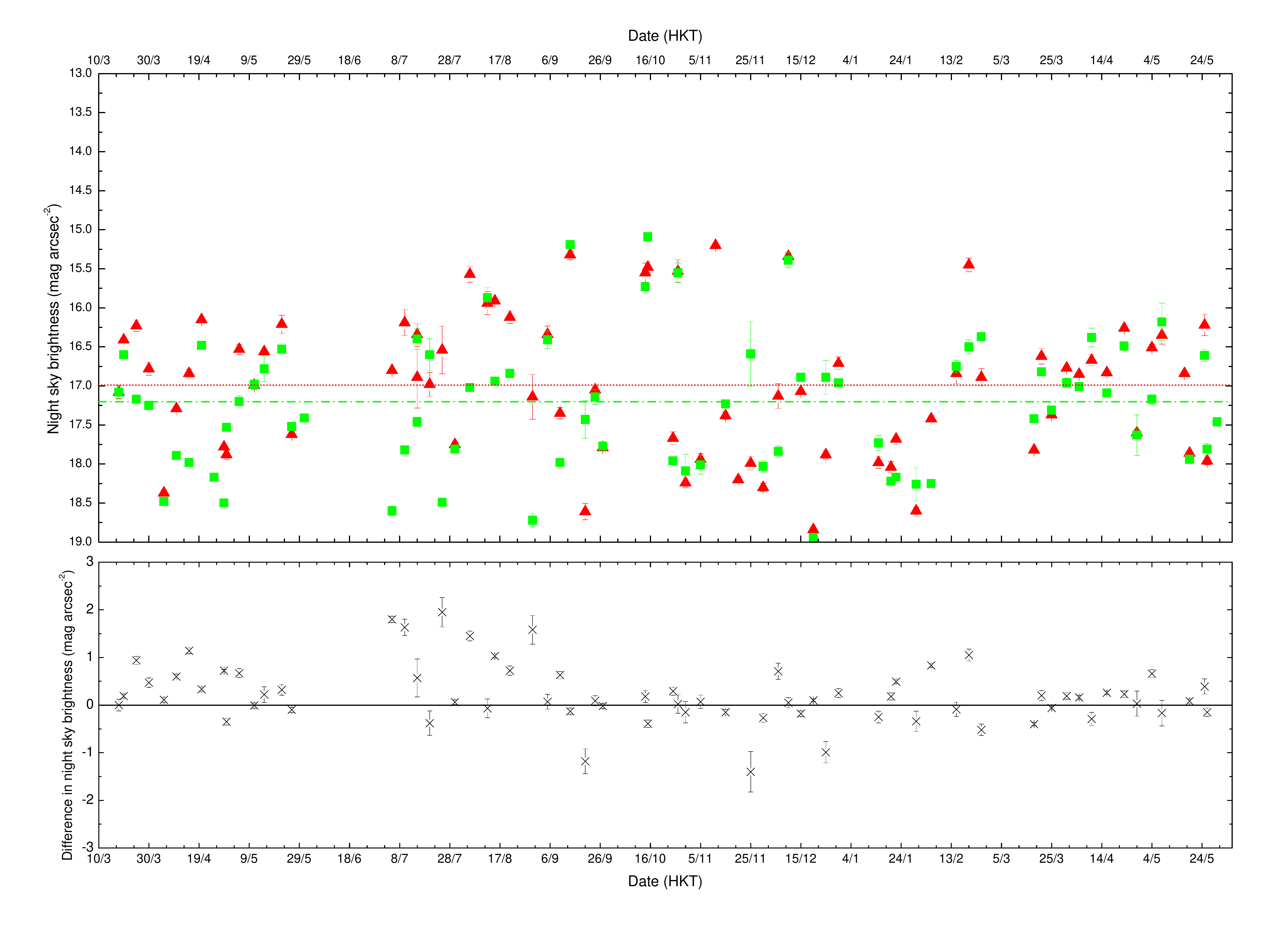}
\caption{Same as Figure~\ref{fig:site35_delta_m} for site 37. }\label{fig:site37_delta_m}
\end{figure}

The long-term variations of NSB over time for sites 35 and 37 are plotted in the upper panels of Figures~\ref{fig:site35_delta_m} and \ref{fig:site37_delta_m} respectively. In these figures, the measured NSB at 9:30pm, 11:30pm, and other times are all  shown, along with the survey-average NSB level of each site measured at 9:30pm (red dotted line) and 11:30pm (green dashed line) for comparison. The NSB on both sites are found to vary by a large amount over the survey period, with the brightest night-sky's NSB value of order 3 to 4 mag~arcsec$^{-2}$ lower than that of darkest sky at that same location and at the same time throughout different days of the year. As discussed above, these observations have been taken under almost identical artificial lighting environments at the same time in each day and are considered to be independent of human influences. 

The observed fluctuations thus are probably be due to non-human factors that are meteorological and/or environmental in nature, such as temperature, humidity, haze condition, and air pollution. Variations of one or more of these factors can affect the observed NSB. A study of the NSB variation with air quality factors will be presented below in Section~\ref{sec:air_VS_NSB}. Other possible causes for the variations observed are astronomical factors such as light from bright stars and planets, of which the most significant would be the variation of moonlight due to positions of the Moon in the sky and varying Moon phase within a month \citep{krisciunas:1991}. On the other hand, the short-term instrumental malfunctions of SQMs (though belived to be reliable over long-term, see Section~\ref{subsec:sqm}) or human errors, as suspected of the very bright night-skies recorded at site 35 at the beginning of the survey seen in the upper panel of Figure~\ref{fig:site35_delta_m}, could not be ruled out.

\subsection{Comparison of light pollution conditions at selected sites} \label{subsec:nsblongdiff}
The large amount of data collected at the selected sites 35 and 37 allows for more detailed studies of the light pollution conditions there. First, the variation of the NSB within the same evening for these two sites were also studied. The measured intranight NSB differences between 11:30pm and 9:30pm for sites 35 and 37 were investigated and plotted in the bottom panels of Figures~\ref{fig:site35_delta_m} and \ref{fig:site37_delta_m} respectively. Similar to trend as found for all other sites in Section~\ref{subsec:timesurvey}, the night-sky was generally darker at 11:30pm (71\% of the time for site 35 and 64\% for site 37) compared to that at 9:30pm, due partly to the turn-off of public lightings at 11:00pm in Hong Kong. 

Second, similar to the snapshot analysis in Section~\ref{subsec:snapshots}, cross-site NSB difference for data taken at the same night at 9:30pm and 11:30pm could be investigated. For the measurements taken at 9:30pm, 82\% (32 out of 39) indicated that the night-sky at site 37 was darker than at site 35, with an average NSB difference of 0.8 mag~arcsec$^{-2}$. For the 11:30pm data, 95\% (40 of 42) indicated that site 37 was darker with an average NSB difference 1.0 mag~arcsec$^{-2}$. 
These results can probably be explained by the environments of the sites: as described in Section~\ref{subsec:samesite} and marked in Figure~\ref{fig:sitedistrib}, site 37 is surrounded by mostly woodland and hills of country parks on the Lantau Island. On the contrary, the suburban town of Yuen Long (populations 427,000 from the 2006 Population By-census data) is only 3.5~km north of site 35. Differences in the built environment and thus the artificial lighting conditions should contribute to the observed NSB difference. Both analyses provide a strong argument that light from outdoor lightings indeed contributes to the brightening of the night-sky. 

Other long-term trends of the observed NSB from data collected at these two sites were studied. Seasonal effects in the night-sky were investigated but no reliable trend in the data sets could be established, except for an apparent of brightening of the night-sky for site 35 after the end of 2009 January. However, this may have been caused by an unreported small change of the external lighting condition near that location. Moreover, the lighting usage pattern in the city was investigated by studying the NSB measured on weekend (Friday to Sunday) and public holidays and their eves versus weekday (Monday to Thursday) evenings, but no significant trend could be observed. 

\subsection{Night-sky brightness variation against air pollutant concentrations} \label{sec:air_VS_NSB}

The presence of air pollutants in the atmosphere, such as soot, aerosol molecules, and other particulates could have effects on the brightness of the night-sky by the strong scattering properties of these atmospheric contaminants. Stray light from the ground directed upwards could be scattered to all directions by these ingredients, thus spreading its effects to a wider region. Air quality data taken by the Environmental Protection Department (EPD) of the Hong Kong government were downloaded\footnote{download site: \url{http://epic.epd.gov.hk/ca/uid/airdata/p/1}} and extracted to compare with the night-sky measurements. Data from the Tung Chung station of the air quality monitoring network of the EPD (location marked with a pink square in Figure~\ref{fig:sitedistrib}) was obtained for comparison with the NSB data collected from the 23 observation sites within 1.5~km away from that EPD station. Tung Chung is a suburban town in southwestern Hong Kong which is adjacent to the Chak Lap Kok Hong Kong International Airport which operates 24 hours a day and NSB measurements there are expected to be affected by the strong lightings from this busy facility.

\begin{figure}
\includegraphics[width=10.5cm]{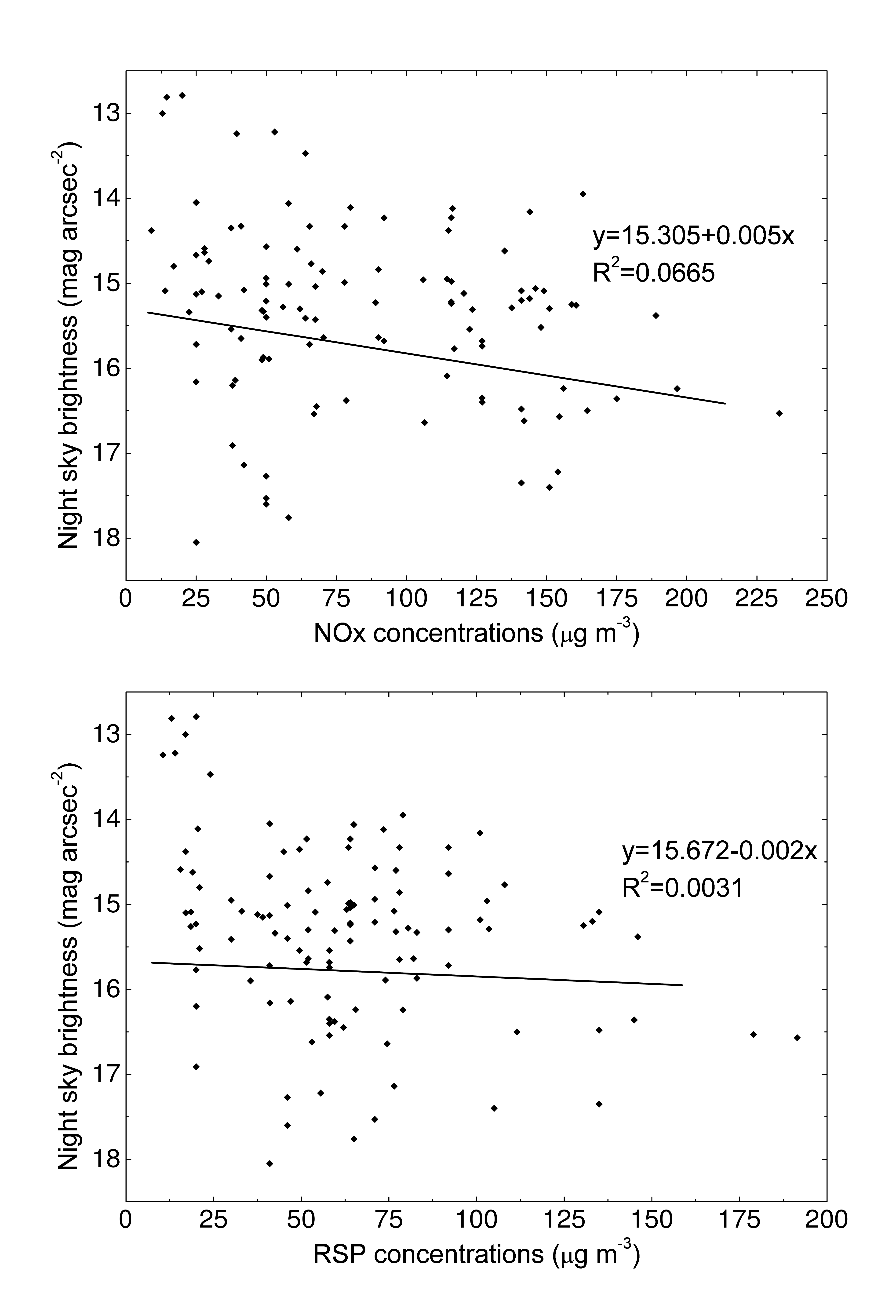}
\caption{Night-Sky Brightness (NSB) taken at 23 observation sites in Tung Chung in units of mag~arcsec$^{-2}$ plotted against the concentrations of nitrogen oxides (NO$_x$) and respirable suspended particulate (RSP) measured from the EPD air pollution observing station in Tung Chung. Small NSB values represent brighter sky while large NSB values represent darker sky. The sampling period covered the entire survey period. The lines indicate the linear best fits with equations and the values of coefficients of determination ($R^2$) marked.}
\label{fig:TC_NSB_vs_air}
\end{figure}

The measured values of NSB were plotted against the concentrations of nitrogen oxides (NO$_x$) and the ambient respirable suspended particulate (RSP) from the air monitoring station in Figure~\ref{fig:TC_NSB_vs_air}. The pollutant concentration data were taken every hour at the hour and the two values before and after the 9:30pm and 11:30pm observing time were averaged for comparison with the NSB measurements. A weak trend ($R^2\sim 0.07$) of dimmer night-skies with increasing concentrations of NO$_x$ were observed while no relation ($R^2\sim 0$) between NSB and concentrations of RSP could be established. It was also clear that a wide range of NSB could be observed for similar values of NO$_x$ and RSP readings, thus indicating that other factors should be contributing to the observed night light conditions. A limit of this kind of study was that the EPD air quality measurement stations were usually installed not high above the ground level (21m above ground on the rooftop of a building for the Tung Chung station) and hence the readings could be strongly affected by the activities in the nearby dense residential area. Therefore the air quality parameters measured might not reflect the quality of the higher atmosphere directly above the zenith. 

\section{Discussions}
\label{sec:discussion}
Light pollution poses negative impacts on our natural and our living environment, robs us of our night-sky, and represents a waste of electric energy and hence contributes to the environmental degradations of the Earth. Previous studies of this problem can be grouped under two main categories: astronomical and remote-sensing. In the first category, precise NSB measurements can be made at research observatories using large astronomical telescopes, though at only a handful locations (usually near research observatories). Wide applications of this technique are limited by the expertise knowledge and dedicated equipment required and thus cannot usually be conducted over a large geographic area. There have also been large campaigns to survey the quality of the night-sky by amateur astronomers using simple visual techniques over a large area. While the scientific accuracy of these programs could be limited by the varying experiences of the participants, these cheap and non-technical activities do provide rough conditions of light pollutions over wide areas and raise awareness of this problem in the public. In the second category, accurate data with high spatial resolution (down to $\sim$~0.5~km) can be analyzed over a large region. However, measurements could be hampered by the presence of cloud in the data and thus analyis can only be made on cloudless nights \citep{chalkias:2006}. Furthermore, the amount of night-time imaging satellite data publicly available is still very low and thus the data available, while complete in geographic coverage, will not be sufficient for either short-term or long-term studies of the evolution of the night-sky condition. A dedicated night-sky monitoring satellite could drastically improve the situation (see for example the \textit{NightSat} satellite proposal by \citealt{elvidge:2007}).

The present survey presents one way to conduct an accurate survey of the quality of the night-sky over a large metropolis. The time of measurements and the duration of the project are selected so that both short-term (intranight) and long-term properties of the night-sky can be studied. The inclusion of the general public in the data collection helps not only to expand the geographic and time coverage of the survey, but also to raise awareness of the light pollution condition in the society. 

While the present survey provides an overview of the condition of night-sky in Hong Kong, a more comprehensive study with this current data set cannot be conducted. First and foremost, by relying on volunteering participants to take and report measurements, strict quality control of the data collected could not be enabled as participants could not be ensured to follow all the guidelines provided. In particular, it was difficult to determine whether all observing sites had wide field-of-views and were not under direct illumination of some lighting sources.  Second, as the survey did not want to overwork the participants, they took NSB readings only every five days at two times (9:30pm and 11:30pm) of the evening. The limited time coverage prohibited detailed studies of the behavior of the NSB within the same evening. Third, as the participation in the survey was on a voluntary basis, the overall data collection efficiency allows long-term monitoring of the NSB only at limited sites. Fourth, as the volunteers were allowed to select NSB observing sites of their choices, there were only a handful of cases when the survey had obtained additional meteorological and air pollutant concentration data for comparison with the NSB measurements. This limits the ability to correlate the various factors which may affect the quality of the night-sky. Finally, as most of the volunteers were students who were free to select the observing sites near their homes (mostly within heavily populated areas), the survey would not able to ensure that observing sites were evenly distributed or every population density level had an equal chance of having an observation. 

The present survey results lend strong support for the development of regulations on the usage of external lightings in Hong Kong to conserve the night-sky and the natural environment. Dark sky friendly outdoor light housings with full cut-off designs \citep{full_cut_off_1} which reflect light totally downward should be required or at least encouraged for use by government and promoted to the general public. Better design of light shielding not only reduces light pollution by avoiding direct uplight effectively, but also means reduction of electrical energy consumption since we can apply light bulbs with lower power while achieving the same light intensity. Moreover, regulations for turn-off time, pointing direction, and timing-pattern (no rapidly flashing lights) of external lightings can also contribute to relieve the problem \citep{massey:2010}. A full understanding of the existing condition is essential for the society in devising effective mechanisms to control the problem of light pollution in Hong Kong. In particular, special focus should be placed on the preservation of the night-sky in rural areas, with ideas such as the establishment of Night-Sky Preservation Regions to be considered (see for example the proposal prepared by a local green group Friends of the Earth Hong Kong: \url{http://www.foe.org.hk/}). The entire environment and the ecosystem, not just astronomers, would be benefited from the success of the battle for reducing light pollution.

\section{Conclusions}
\label{sec:conclusion}
A survey was conducted to accurately monitor the light pollution condition in Hong Kong. Night-sky brightness (NSB) data were collected for a 15-month period from 2008 March to 2009 May at 199 distinct urban and rural locations widely distributed throughout the city. Non-specialist volunteers participating in the survey used a small and easy-to-use device called the Sky Quality Meter (SQM) to take instantaneous readings of the NSB and reported the data online through the survey webpage. 
With the 1,957 NSB data sets collected, a Hong Kong Light Pollution Map showing the distribution of the light pollution conditions in Hong Kong was created. It shows that the light pollution in Hong Kong is severe~--- the overall average NSB reading obtained in this survey is 16.1 mag~arcsec$^{-2}$, or 160 times brighter in flux than the dark site recommendation defined by the International Astronomical Union (NSB level of 21.6 mag~arcsec$^{-2}$ \citep{smith:1979}).

Artificial lightings in dense residential and commercial urban area contribute strongly to the problem of light pollution, as reflected by the urban night-skies being on average 100 times brighter than the darkest rural sites. This huge gap provides an evidence to support for the development of regulations on the usage of external lightings in Hong Kong. Trends of the observed NSB with population densities and land usage both also indicated that human activities strongly affect the brightness of the night-sky. Time is also an important determining factor of the night-sky darkness. The sky after 11pm (Hong Kong local time, same as below) was found to be on average darker than the sky before 11pm, attributed possibly to the turning off of a majority of public and private outdoor lighting throughout the city in late evenings. On the other hand, the large long-term amplitude variation of the NSB at two selected sites indicate that other non-human factors such as astronomical, meteorological, and/or environmental should also contribute to the observed NSB, even though no concrete relation between the observed sky brightness and air pollutant concentrations could be established in a limited study presented.



. 

The University of Hong Kong is now extending the survey by constructing a network of observing stations in which the night-sky conditions will be simultaneously and automatically monitored at multiple sites in Hong Kong at a high frequency for a long period of time. A few locations from this survey were selected for this detailed study that is currently underway.  Therefore data from this study will be part of a larger data base in the future. 
This project is expected to be concluded in 2012 and should provide a more comprehensive picture of the light pollution situation in the city.  \\

\textbf{\Large{Acknowledgements}}\\

We acknowledge support of the Environment and Conservation Fund (ECF) of the Environmental Protection Department of the Hong Kong Special Administrative Region (HKSAR) Government, and the Woo Wheelock Green Fund for the project titled \textit{A Survey of Light Pollution in Hong Kong} (Project No.: 2007-01). Co-organizers of the project include the Hong Kong Space Museum, the Sky Observers' Association (Hong Kong), Ho Koon Nature Education cum Astronomical Centre (Sponsored by Sik Sik Yuen), and The Camping Association of Hong Kong, China, Ltd. We would like to thank all the 171 volunteering participants who have contributed data to our project. The list of organizations can be found at \url{http://nightsky.physics.hku.hk/webcoorg.asp}. Data and technical supports from the HKSAR government's Census and Statistics Department, Environmental Protection Department, Lands Department, Planning Department, and Registration and Electoral Office are also acknowledged.

\bibliographystyle{aa}
\bibliography{main-astro-ph2-p7}

\end{document}